\documentstyle[12pt]{article}

\renewcommand{\thefootnote}{\fnsymbol{footnote}}

\newlength{\extraspace}
\newlength{\extraspaces}
\newcommand{\beq}{\begin{equation}}
\newcommand{\eeq}{\end{equation}}
\newcommand{\bea}{\begin{eqnarray}}
\newcommand{\eea}{\end{eqnarray}}

\newcommand{\goto}{\rightarrow}

\newcommand{\zb}{\bar{z}}

\newcommand{\hb}{\bar{h}}

\newcommand{\al}{\alpha}

\newcommand{\xb}{\bar{x}}

\newcommand{\tPhi}{\tilde{\Phi}}
\begin{document}
\addtolength{\baselineskip}{.8mm}
\oddsidemargin 5mm
\renewcommand{\thefootnote}{\fnsymbol{footnote}}
\newpage
\setcounter{page}{0}
\begin{titlepage}
\begin{flushright} 
{\tt hep-th/0009096
}
 \end{flushright}
\bigskip
\bigskip
 
\begin{center}
{\Large Logarithmic CFT on the Boundary and the World-Sheet}
\bigskip 
\bigskip

{Alex Lewis\footnote{alex@thphys.may.ie\\supported by Enterprise Ireland
grant no. SC/98/739}\\}  \bigskip
{Department of Mathematical Physics, National University of Ireland,
Maynooth, Republic of Ireland.}

\end{center}
\begin{center}
\footnotesize
        
	\end{center}               

\normalsize 
\bigskip 
\bigskip
\begin{center}
			{\bf Abstract}
\end{center}
{The correspondences between logarithmic operators in the CFTs on the
boundary of $AdS_3$ and on the world-sheet and dipole fields in the bulk
are studied using the free field formulation of the $SL(2,C)/SU(2)$ WZNW
model.  We find that logarithmic operators on the boundary are related
to operators on the world-sheet which are in indecomposable
representations of SL(2). The Knizhnik-Zamolodchikov equation is used to
determine the conditions for those representations to appear in the
operator product expansions of the model. }

 \end{titlepage} 
\newpage


\section{Introduction}

String theory on $AdS_3$ is described by the $SL(2,R)$ WZNW model (or its
Euclidean version, the $SL(2,C)/SU(2)$ coset model). This
theory is interesting for a number of reasons. It provides a simple
example   of a string theory in a non-trivial background, and it is also
one of the simplest examples of a non-rational CFT \cite{ref1,gaw}. It
is closely related to string theory in black hole backgrounds such as
the BTZ black hole \cite{btz} and the $SL(2,R)/U(1)$ coset model which
describes the two dimensional black hole \cite{wittenbh}.
There has
been a great deal of interest in the model recently because it is also
the simplest example of the $AdS$/CFT correspondence 
\cite{mal}--\cite{ios}. The model also has applications in condensed
matter physics, to the theory of disordered systems \cite{ctt} and the
plateau transition in the quantum Hall effect \cite{bkstt}.

In addition, the $SL(2,R)$ WZNW model 
 is thought to be an example of a Logarithmic CFT
\cite{gur}. The first evidence of this  was seen in \cite{bk}, where an
asymptotic solution for a correlation function of a gravitationally
dressed CFT was found which had a logarithmic singularity, indicating
that there are logarithmic operators in the theory. Two dimensional gravity
has conserved currents which have the same Kac-Moody algebra as the
currents in the $SL(2,R)$ WZNW model  \cite{kpz}, and the Ward identities
which were used to find correlation functions in that model are equivalent
to the Knizhnik-Zamolodchikov equations for correlation functions of the
WZNW model. Thus, the logarithms that were found in some four-point
functions  in  2-dimensional gravity should also appear in the WZNW model
\cite{kls}. Logarithmic behaviour has also been found  in correlation
functions of fields in finite dimensional representations of $SL(2)$ in the
$SL(2,C)/SU(2)$ coset model \cite{ctt},  and recently   in exact
solutions of the Knizhnik-Zamolodchikov equations for correlation functions
of fields in infinite dimensional representations of $SL(2)$ as well
\cite{kt,ns}.

The presence of logarithmic operators in the spectrum of the $SL(2,R)$
model raises several questions. The existence of a logarithmic operator
always implies that the primary operator which is part of the same
indecomposable representation of the Virasoro algebra has zero norm,
 so if all the ghosts are to be eliminated for the theory by
the Virasoro constraints, we should not have any logarithmic operators.
The no-ghost theorem for $AdS_3$ restricts the  fields that can appear in
the string theory \cite{egp}, and there may be a further restriction
if the spectral flow is a symmetry of the theory \cite{mo1}. It is
therefore possible that when these restrictions are imposed, the fields
which have logarithmic correlation functions are excluded, and there are
no logarithmic operators. This is exactly the situation that occurs in the
minimal models, or in WZNW models on compact groups -- there are operators
which have logarithmic correlation functions, but they are always outside
the set of operators which occur in a unitary or minimal model.
However, world-sheet logarithmic operators are
thought to generate  zero modes in target space which restore
symmetries that are broken by the string background \cite{km}. This has
mainly  been used to study D-brane recoil (see eg. \cite{recoil}), but
logarithmic operators are expected to occur more generally in string
theories in non-trivial backgrounds. One of the examples of a string
theory with logarithmic operators discussed in \cite{km} is the 2D black
hole which is closely related to the $SL(2,R)$ WZNW model.   From this
point of view, it would not be very surprising if string theory on $AdS_3$
also has logarithmic operators.

There are two CFTs associated with the string theory on $AdS_3$ - the
world-sheet theory and the dual CFT on the boundary of $AdS_3$ that is
related to the string theory by the AdS/CFT correspondence. The dual theory
can also be a LCFT, if there are singletons in the
bulk of  $AdS_3$ \cite{kogan,gka}. If singletons in $AdS_3$ are related to
logarithmic operators on the world sheet, we will therefore have a duality
between two LCFTs. 

In the next section we briefly review some basic facts about LCFT, the
relation between singletons and logarithmic operators in the $AdS_3$/CFT
correspondence, and the logarithms in correlation functions of the
$SL(2,R)$ WZNW model.  In section (3) we use the free field formulation of
the WZNW model  to construct logarithmic operators, and determine how
logarithmic operators on the world-sheet and on the boundary of $AdS_3$
and fields in the bulk are related. In section (4) we use the
Knizhnik-Zamolodchikov equations to investigate the conformal blocks and
OPE of the WZNW model, and determine which OPEs include logarithmic
operators.

\section{Logarithmic CFT}

A LCFT differs from an ordinary CFT in that the Virasoro generator
$L_0$ is not diagonalizable. In addition to the usual primary and
descendant fields, it includes pairs of operators which form Jordan
blocks for $L_0$:
\beq
L_0C = hC,~~~~~~~L_0D = hD +C
\label{jordan}\eeq
The fields  $C$ and $D$ are therefore in a reducible, but
indecomposable representation of the Virasoro algebra. This type of
operator was first introduced in \cite{gur}, and since then the theory of
LCFTs has been developed in many papers, including \cite{lcft}.
 The operators $C$, $D$ have the two-point functions 
\bea \langle
C(z_1,\zb_1)C(z_2,\zb_2) \rangle &=&0 \nonumber \\ \langle
D(z_1,\zb_1)C(z_2,\zb_2)\rangle &=& \frac{c}{|z_1-z_2|^{4h}} \nonumber \\
\langle D(z_1,\zb_1)D(z_2,\zb_2)\rangle &=&
\frac{1}{|z_1-z_2|^{4h}}(d-2c\ln|z_1-z_2|^2)
\label{CD} \eea
Here $c$ is determined by the normalisation of $C$ and $D$ and $d$ is
arbitrary -- it can be set to any value, using the symmetry of the theory
under $D \goto D + \lambda C$, which leaves (\ref{jordan}) unchanged. In
the next section we will use the free field formulation of the WZNW model
on $SL(2,C)/(SU(2)$ to identify the logarithmic operators in the
$AdS_3/CFT$ correspondence, so we begin by reviewing the free field 
formulation of LCFTs with $c<1$, which was developed in \cite{kl}. In that
case,  there is a single free field, $\phi(z,\zb)$, and the stress tensor
is  \beq 
T(z)=-\frac{1}{4}\partial_z\phi\partial_{\bar{z}}\phi\ +iQ\partial^2_z\phi
 \eeq 
for a central charge $c=1-24Q^2$. The primary field are
the exponentials of $\phi$: $C_\al = e^{i\al\phi}$ with  conformal weights
$h_\al = \hb_\al = \al(\al-2Q)$.  Logarithmic operators can be represented
as derivatives of ordinary fields with respect to $h$: 
\beq
D_\al  = \frac{d}{dh} C_\al =\frac{dh}{d\al}\phi e^{i\al\phi} =
\frac{i}{\al-Q}\phi e^{i\al\phi}
\eeq
Then $D_\al$ and $C_\al$ form the Jordan block for $L_0$ as in
(\ref{jordan}). However we cannot write down an operator of this form when
$\al=Q$, so that $h=-Q^2$ and $\frac{dh}{d\al}=0$. In that case, the field
$\phi e^{i\al\phi}$ is a primary field -- in Liouville theory it is the
puncture operator. However, although there are no logarithms in the two
point functions, there are logarithms in the four point functions of the
field with this dimension. This indicates that logarithmic operators must
appear in the operator product expansion (OPE) of the field with itself,
and so the full spectrum of the theory must include logarithmic operators
if it includes the primary operator with $\al=Q$. In an ordinary CFT the
OPE of two primary fields has the form 
\beq O_1(z_1)O_2(z_2) =
\sum_i\frac{f^i_{12}}{(z_1-z_2)^{h_1+h_2-h-i}}O_i(z_2) + \cdots 
\eeq
Where the $(\cdots)$ indicates the contributions of descendant fields.
An OPE of this form implies an s-channel expansion of four point
functions as
\beq
\langle O_i(z_1)O_j(z_2)O_j(z_3)O_i(z_4) \rangle =
\sum_{kl}
\frac{(f^k_{ij})^2}{|z_{12}|^{h_i+h_j}|x_{34}|^{h_i+h_j}}
F^k_{ij}(z)F^k_{ij}(\zb) 
\eeq
where $z=z_{12}z_{34}/z_{13}z_{24}$ and $F^k_{ij}(z)$ has an expansion in
powers of $z$ around the point $z=0$. In a LCFT, if one of the primary
fields $O_i$ is the field with dimension $-Q^2/2$, some of the conformal
blocks instead have $F^k_{ij} \sim z^{2h_k}(1 + \ln z + \dots)$, which
requires an OPE of the form 
\beq O_i(z_1)O_j(z_2) \sim 
\frac{1}{|z_1-z_2|^{h_i+h_j-h}}(D + C \ln |z_1-z_2|^2) + \cdots
\label{ope2}\eeq
For that reason the field which has the form $\phi
e^{\al\phi}$ but is not a logarithmic operator was called a
pre-logarithmic operator in \cite{kl}. In the minimal models, with
central charge $c_{p,q} = 1-6(p-q)^2/pq$, the field with dimension
$\al=Q$ is the degenerate primary field with dimension $h_{p,q}$ at 
the
corner of the Kac table. It is therefore excluded from the spectrum of the
minimal models, which consists of operators with dimensions $h_{r,s}$, $1
\leq r \leq p-1$, $1 \leq s \leq q-1$. However, it is possible to define
expanded models with the same central charge as the minimal models, which
include field with conformal weights from the edge of the Kac table, and
these models do have logarithmic operators.  To determine whether
logarithmic operators can consistently be excluded from the spectrum of
the $SL(2,R)$ WZNW model, we  therefore need
to find out which primary fields will have logarithms in four point
functions in that theory.. 

\subsection{LCFT and Singletons}

Since logarithmic operators are in  indecomposable representations of the
conformal algebra, in the AdS/CFT correspondence we can expect them to be
related to fields which form similar indecomposable representations in the
bulk of AdS. It was observed in \cite{kogan} that singletons  in
AdS are objects which are in just that kind of representation
\cite{singleton}. A free singleton theory can be formulated in terms of a
massive dipole field $A$ which satisfies 
\beq
(\nabla^2-m^2)^2A = 0
\label{dipole1}\eeq
When $m^2=-D^2/4$, which is the lower bound for a stable massive scalar
field on $AdS_{D+1}$, this equation has a singleton solution \cite{ff}. Eq.
(\ref{dipole1}) can be solved by introducing a second field B with the
equations of motion \bea
(\nabla^2-m^2)B = 0 \nonumber \\
(\nabla^2-m^2)A - \mu^2B = 0
\label{dipole2}\eea
The action for field in $AdS_{D+1}$ with these equations of motion is
\beq
S = \int d^{D+1}x \sqrt{g}\left(g^{\mu\nu}\partial_{\mu}A\partial_{\nu}B  -
 m^2 AB - \frac{\mu^2}{2} B^2 \right)
\label{dipoleaction}\eeq
Correlation functions of operators $O_i(\vec{x})$ in the CFT on the
boundary of $AdS_{D+1}$ which correspond to fields $\Phi_i$ in the bulk can
be calculated from the bulk action  using the relation  \cite{gkp,witten}
\beq
\left< \exp\left[ \sum_{i}\int~ d^Dx ~ 
\lambda_i (\vec{x}) O_i(\vec{x}_i)\right]\right>_{CFT} =  
e^{- S[\{\Phi_i\}]}\arrowvert_{\Phi_i(\delta AdS)  = \lambda_i(\vec{x})}
\label{adscft}\eeq
In \cite{gka} and \cite{kogan}, this relation was used to find the
two-point functions of operators on the boundary which correspond to the
dipole fields with the action (\ref{dipoleaction}) in the bulk. The result
is that for $m^2 \neq -D^2/4$, the two point functions are precisely those
given by eq. (\ref{CD}), with $h$ given by $m^2 = 2h(2h-D)$, indicating
that dipole fields in the bulk lead to logarithmic operators with those
dimensions on the boundary. In \cite{kogan}, but not in \cite{gka}, it was
found that the logarithmic correlation functions vanish for $m^2 =
-D^2/4$; we will return to this discrepancy in section (3.2).

\subsection{Four-Point Functions in the WZNW model on $SL(2,R)$ or
$SL(2,C)/SU(2)$}

String theory on $AdS_3$ is described by the WZNW model on $SL(2,R)$, or
its Euclidean version $SL(2,C)/SU(2)$. This model has the
$\widehat{SL(2)}\times\widehat{SL(2)}$ algebra generated by left- and
right-moving currents with the OPEs
\bea
J^3(z_1)J^3(z_2) &\sim& -\frac{k}{2(z_1-z_2)^2} \nonumber \\
J^3(z_1)J^\pm(z_2) &\sim& \pm\frac{J^\pm(z_2)}{z_1-z_2} \nonumber \\
J^-(z_1)J^+(z_2) &\sim& \frac{k}{(z_1-z_2)^2} + \frac{2J^3(z_2)}{z_1-z_2} 
\eea
and similar relations for $\bar{J}^a(\zb)$. A primary operator
$\Phi_j(z,\zb)$ in the representation of $SL(2)$ labeled by $j$ is
defined by the OPE
\beq
J^a(z_1)\Phi_j(z_2) \sim \frac{D^a\Phi_j}{z_1-z_2}
\eeq
where $D^a$ are generators of $SL(2)$. It is useful to use the
representation of the generators introduced in \cite{zf}
\beq
D^- = -\partial_x,~~~~~D^0= x\partial_x+j,~~~~~ D^+ = -x^2\partial_x-2jx
\label{D0pm}\eeq
so that the fields are now functions of two complex variables $(x,\xb)$
and $(z,\zb)$. The stress tensor of the model is given by the Sugarawa
construction
\beq
T(z) = \frac{1}{k-2}:J^a(z)J^a(z): = \frac{1}{k-2} :\left[\frac12 J^+J^- +
\frac12 J^-J^+ -J^0J^0\right]:
\eeq
So  the primary fields have conformal dimensions
$h_j=\bar{h}_j=j(1-j)/(k-2)$.
The spectrum of the WZNW model includes operators in
representations from both the principle continuous series of $SL(2)$ with
$j=\frac12+is$, with $s$ real, and the principle discrete representations
with $j$ real. There are also other representations which are obtained
from these by the spectral flow \cite{mo1}, but we will not be interested
in those representations in this paper.  We will concentrate on operators
with real $j>0$, which correspond to local fields on the boundary of
$AdS_3$.   For normalizable functions on $AdS_3$, we should have
$j>1/2$, and for all negative-norm states to be removed by the
no-ghost theorem, we need $j<k/2$ \cite{egp}. If the spectral flow is a
symmetry of theory, the spectrum is truncated further, and we have
$1/2<j<(k-1)/2$ \cite{mo1}. When $j$ is equal to the upper or lower bound,
a continuous representation appears ($j=1/2$ is part of the principle
continuous series, and $j-(k-1)/2$ is obtained from $j=1/2$ under 
the spectral flow). To determine if the spectrum also includes
logarithmic operators, the question we need to ask is therefore, are there
logarithms in the four point functions (or OPEs) of primary fields with
$1/2 \leq j \leq (k-1)/2$?

Correlation functions of the WZNW model 
satisfy the Knizhnik-Zamolodchikov (KZ) equation
\beq
\left[ (k-2)\frac{\partial}{\partial z_i} + \sum_{j\neq i}
\frac{D^a_iD^a_j}{z_i-z_j}\right] \langle \Phi_1(z_1)\cdots\Phi_n(z_n)
\rangle =0
\eeq
For the WZNW on a compact group, the KZ equation reduces to a system of
ordinary differential equations. For $SL(2,R)$, using the representations
(\ref{D0pm}), we instead get a partial differential equation. The four
point function can be expressed in terms of the cross ratios
$z=z_{12}z_{34}/z_{13}z_{24}$ and $x=x_{12}x_{34}/x_{13}x_{24}$, and
$z_{ij}=z_i-z_j$, as
\bea
\langle \Phi_{j_1}(x_1,z_1)\cdots\Phi_{j_4}(x_4,z_4)\rangle =
|x_{24}^{-2j_2}x_{14}^{\beta_1}x_{13}^{\beta_2}x_{34}^{\beta_3}
z_{24}^{-2h_2}x_{14}^{\gamma_1}z_{13}^{\gamma_2}z_{34}^{\gamma_3}
|^2 G(x,\xb,z,\zb) \label{4point}\\
\beta_1=j_2+j_3-j_4-j_1,~~~~ \beta_2=j_4-j_1-j_2-j_3,~~~~~
\beta_3=j_2+j_1-j_4-j_3 \nonumber\\
\gamma_1=h_2+h_3-h_4-h_1,~~~~ \gamma_2=h_4-h_1-h_2-h_3,~~~~~
\gamma_3=h_2+h_1-h_4-h_3 \nonumber
\eea
$G(x,\xb,z,\zb)$ can be factorised into the conformal blocks, as
\beq
G(x,\xb,z,\zb) = \sum_{ij}U_{ij}F_i(x,z)F_i(\xb,\zb)
\label{blocks}\eeq
 and the KZ equations for the conformal blocks are then \cite{zf,teschner}
\beq
(k-2)\frac{\partial}{\partial z} F(x,z) = \left[\frac{\cal P}{z} +
\frac{\cal Q}{z-1} \right] F(x,z)
\label{KZ}\eeq
where
\bea
{\cal P} &=& -x^2(1-x)\frac{\partial^2}{\partial x^2} + \left[ (\tau +
1)x^2 - 2j_1x-2j_2x(1-x) \right] \frac{\partial}{\partial x} +2j_2\tau x -
2j_1j_2 \nonumber \\
{\cal Q} &=& -x(1-x)^2\frac{\partial^2}{\partial x^2} - \left[ (\tau +
1)(1-x)^2 - 2j_3(1-x)-2j_2x(1-x) \right] \frac{\partial}{\partial x}
\label{PQ}\\
&&+2j_2\tau (1-x) - 2j_3j_2 \nonumber
\eea
and $\tau=j_1+j_2+j_3-j_4$. To find the complete four-point function, and
the OPE of the operators in it, we need to determine which conformal
blocks are included in a solution which satisfies the conditions of
crossing symmetry and is single valued in the region of the singularities
at $x=0,1,\infty$. The complete solutions for some four point functions,
with logarithmic behaviour which requires logarithmic operators in the
OPE, were found in \cite{ns,kt}, but all the correlation functions
calculated in which logarithms were found included operators with $j<1/2$.
For example if $j_1=j_3$ and $j_2=j_4=0$, the solution is \cite{ns}  
\beq
G(x,\xb,z,\zb) = \ln\left|\frac{1-x}{x}\right| + \frac{2j_1-1}{k-2}  
\ln\left|\frac{1-z}{z}\right|
\eeq
This correlation function involves the operator in a non-trivial
representation with $j=0$, which is not expected to be part of the
spectrum of string theory on $AdS_3$, although it may be relevant to the
theory for the Quantum Hall plateau transition \cite{bkstt,kt}. However,
the KZ equation has a symmetry which can be used to relate this four point
function to the four point function with $j_2=j_4=1$. If $1-2j_2$ is a
non-negative integer, then
 $\frac{\partial^{1-2j_2}}{\partial x^{1-2j_2}}F(x,z)$
is a solution to the KZ equation with $j_2$ replaced by $1-j_2$ \cite{ns}.
Using this symmetry we  can see that the four point function with $j_2=1$,
$j_1=j_3$, $ j_4=0$ has no logarithms -- it is simply
\beq
G(x,\xb,z,\zb) = -\frac{1}{x} - \frac{1}{1-x} -\frac{1}{\xb} -
\frac{1}{1-\xb}
\eeq
Therefore the correlation functions calculated in \cite{kt,ns} do not
prove that there are logarithmic operators in the WZNW model. We will look
more closely at correlation function of operators with $j \geq 1/2$ in
section (4).

\section{Logarithmic Operators in $AdS_3$}

 \subsection{Semiclassical Approach}

The action for the WZNW model on $SL(2,C)/SU(2)$ can be written using the
Gauss parameterization of $SL(2,C)$ as \cite{gaw} 
\beq
S = k\int d^2z\left[ \partial\phi\bar\partial\phi +
\bar\partial\gamma\partial\bar\gamma e^{2\phi} \right]
\eeq
This action describes strings propagating in the Euclidean $AdS_3$ target
space, with $(\phi,\gamma,\bar\gamma)$ being coordinates on $AdS_3$. 
The
zero modes of the currents are realised by
\beq
J^-_0 = \partial_\gamma,~~~~~J^0_0 = \gamma\partial_\gamma -
\frac{1}{2}\partial_\phi,~~~~~J^+_0=
\gamma^2\partial_\gamma-\gamma\partial_\phi -
e^{-2\phi}\partial_{\bar\gamma}
\label{J0}\eeq
with similar expressions for $\bar{J}^a$. Primary fields $\Phi_j(z)$
satisfy
  \beq
J^a_0\Phi_j = D^a\Phi_j,~~~~~~~J^a_n\Phi_j=0,~n>0
\label{primary}\eeq
with $D^a$ given by (\ref{D0pm}). The primary fields which satisfy
(\ref{primary}) are \cite{teschner} 
\beq
\Phi_j = \left[ \frac{1}{(\gamma - x)(\bar\gamma - \xb)e^\phi + e^{-\phi}}
\right]^{2j}
\label{vertex}\eeq
These fields can be expanded as
\beq
\Phi_j(x,\xb,z,\zb) = \sum_m x^{-j+m}\xb^{-j+\bar{m}}\Phi_j^m(z,\zb)
\label{xseries}\eeq
The vertex operator (\ref{vertex}) has the form of a boundary-bulk
Green function for a scalar field on $AdS_3$, with mass given by
\beq
m^2 = 4j(1-j)
\label{m2}\eeq
 This is
because $\Phi_j$ is an eigenfunction the Laplacian on $AdS_3$, which in the
$(\phi,\gamma,\bar\gamma)$ coordinates is just $\nabla^2
=2(J^a_0J^a_0+\bar{J}^a_0\bar{J}^a_0)$, with the $J^a_0$ given by eq.
(\ref{J0}). The new variables $x,\xb$ can therefore be considered as
coordinates on the boundary of $AdS_3$ \cite{bort}

As in the $c_{p,q}$ models, we can expect logarithmic fields to be given
by the derivative with respect to $h_j$ of the primary fields. We therefore
define
\beq
\tPhi_j = \frac{d}{dj}\Phi_j  =
\left[\frac{1}{(\gamma - x)(\bar\gamma - \xb)e^\phi + e^{-\phi}}
\right]^{2j} \ln 
\left[ (\gamma - x)(\bar\gamma - \xb)e^\phi + e^{-\phi}
\right]^2
 \eeq
so that $\Phi_j$ and $\tPhi_j$ satisfy
\beq
L_0\Phi_j = h_j\Phi_j,~~~~~~
L_0 \tPhi_j = h_j\tPhi_j + \frac{1-2j}{k-2}\Phi_j
\label{logvertex}\eeq
This has the same form of (\ref{jordan}), with $C=\Phi_j$ and $D =
\frac{k-2}{1-2j}\tPhi_j$ giving the conventional normalization of the
logarithmic pair.  We can also see that when $j=1/2$, $L_0$ becomes
diagonal and so $\tPhi_\frac{1}{2}$ is not a logarithmic operator. 
The action of the currents on $\tPhi_j$ is
 \bea
J^-_0 \tPhi_j &=& D^-\tPhi_j \nonumber \\
J^0_0 \tPhi_j &=& D^0\tPhi_j  + \Phi_j \nonumber \\
J^+_0 \tPhi_j &=& D^+\tPhi_j + 2x\Phi_j
\label{sl2jordan}\eea
where $D^0, D^\pm$ are given be eq. (\ref{D0pm}); also $J^a_n\tPhi_j=0$
for $n \geq 1$. The operators $\tPhi_j$ and $\Phi_j$ are therefore in
an indecomposable representation $SL(2)$ as well as of the Virasoro
algebra. However, these representations are not the same as the
indecomposable representations of the Kac-Moody algebra which were
discussed in \cite{kls,kl}. The difference is that in those
representations, the Casimir operator $J^a_0J^a_0$ was diagonal, and so
the operators in those representations were not necessarily logarithmic
operators as far as the Virasoro algebra was concerned. However, we will
see that  the type of four point functions which were analyzed in
\cite{kls} are consistent with OPEs which include the $\tPhi_j$ type of
field. Eq. (\ref{sl2jordan}) also applies when $j=1/2$, so that
$\tPhi_\frac12$ is in an indecomposable representation of $SL(2)$ even
though it is an ordinary  primary field in an irreducible representation
of the Virasoro algebra.

Eq. (\ref{sl2jordan}) is consistent with eq. (\ref{logvertex}), since
$J^a_0J^a_0\tPhi_j = j(1-j)\tPhi_j +(1-2j)\Phi_j$. Because $J^a_0J^a_0$ is
the Laplacian on $AdS_3$, this means that $\tPhi_j$ is the
boundary-bulk Greens function for  dipole fields with the equation of
motion (\ref{dipole2}), just as the usual vertex
operators $\Phi_j$ are boundary-bulk Greens functions for scalar fields.
As before the mass is given by (\ref{m2}). In fact this follows
immediately from the way we defined $\tPhi_j$, since in \cite{gka,kogan}
this Green function was already found to be given by  differentiating the
Green function for scalar fields with respect to $m^2$ or $j$.  We can
therefore identify  dipole fields in the bulk with vertex operators on the
world sheet which are in indecomposable representations of the global
$SL(2)$ symmetry. To see directly that these vertex operators lead to
logarithmic operators in the CFT on the boundary, we need to show that the
correlation functions of these operators have logarithmic dependence on $x$
as well as $z$.

We can use eq. (\ref{sl2jordan}) to derive Ward identities for correlation
functions of the fields $\tPhi_j$. These are just the same as the Ward
identities for the logarithmic pair of operators in an ordinary LCFT,
 with $z$ replaced by $x$ and $h$ replaced by $j$:
\bea
\left[D^a_{(1})+D^a_{(2}\right]\langle\Phi_j(x_1,z_1)\Phi_j(x_2,z_2)
\rangle &=& 0,~~~~~a=\pm,0\nonumber \\
\left[D^-_{(1)} + D^-_{(2)} \right]\langle\tPhi_j(x_1,z_1)\Phi_j(x_2,z_2) \rangle
&=& \left[D^-_{(1)} + D^-_{(2)}\right] \langle\tPhi_j(x_1,z_1)\tPhi_j(x_2,z_2)
\rangle = 0 \nonumber \\
\left[D^0_{(1)} + D^0_{(2)}\right]  \langle\tPhi_j(x_1,z_1)\Phi_j(x_2,z_2)
\rangle &=& -\langle\Phi_j(x_1,z_1)\Phi_j(x_2,z_2) \rangle \nonumber \\
\left[D^+_{(1)} + D^+_{(2)}\right]  \langle\tPhi_j(x_1,z_1)\Phi_j(x_2,z_2)
\rangle &=& -2x_1\langle\Phi_j(x_1,z_1)\Phi_j(x_2,z_2) \rangle
\label{WardI}\\ 
\left[D^0_{(1)} + D^0_{(2)} \right]
\langle\tPhi_j(x_1,z_1)\tPhi_j(x_2,z_2) \rangle &=&
-\langle\tPhi_j(x_1,z_1)\Phi_j(x_2,z_2) \rangle - 
\langle\Phi_j(x_1,z_1)\tPhi_j(x_2,z_2) \rangle \nonumber \\ 
\left[D^+_{(1)} +
D^+_{(2)}\right]  \langle\tPhi_j(x_1,z_1)\tPhi_j(x_2,z_2) \rangle &=& -2x_1
\langle\Phi_j(x_1,z_1)\tPhi_j(x_2,z_2) \rangle
-2x_2\langle\tPhi_j(x_1,z_1)\Phi_j(x_2,z_2) \rangle 
 \nonumber   \eea
Where $D^a_{(x_i)}$ is given by (\ref{D0pm}) with $x=x_i$, and 
there are
similar equations for the $\xb$ dependence. The conformal  Ward
identities lead to two point functions of the form (\ref{CD}) for
$C=\Phi_j$ and $D=\frac{k-2}{1-2j}\tPhi_j$, except that $c$ and $d$ in
(\ref{CD})  can now depend on $x$.  The solution to eq. (\ref{WardI})
for the two point functions are then 
\bea
\langle \Phi_j(x_1,z_1)\Phi_j(x_2,z_2) \rangle &=& 0 \nonumber \\
\langle \tPhi_j(x_1,z_1)\Phi_j(x_2,z_2) \rangle &=&
\frac{c'}{|z_1-z_2|^{4h_j}|x_1-x_2|^{4j}} \\
\langle \tPhi_j(x_1,z_1)\tPhi_j(x_2,z_2) \rangle &=&
\frac{c'}{|z_1-z_2|^{4h_j}|x_1-x_2|^{4j}}
\left[d'-2\frac{1-2j}{k-2}\ln|z_1-z_2|^2-2\ln|x_1-x_2|^2\right] \nonumber
\label{logxz}\eea
since $(x,\xb)$ are the coordinates on the boundary of $AdS_3$, this
confirms that these fields also give logarithms in the CFT on the
boundary. Of course, to find the complete correlation function is the
string theory we would have to include the contribution from the internal
CFT, and integrate over $z_i$, but that cannot change the way the
correlation functions depend on $x$. The  correspondence we have is
therefore

$
\tPhi_j(\mbox{world-sheet CFT}) \longrightarrow \mbox{dipole fields in
$AdS_3$} \longrightarrow \mbox{LCFT on boundary of $AdS_3$}
$

\subsection{$j=\frac{1}{2}$}

When $j=1/2$, the $\ln z$ term in the two point function vanishes, and
$L_0$ becomes diagonalizable in eq. (\ref{logvertex}). Thus
$\tPhi_\frac{1}{2}$, like the puncture operator in Liouville theory, has
the form of a logarithmic operator but is actually a primary operator.
However, it is still in an undecomposable representation of the Kac-Moody
algebra, and thus we can expect that other indecomposable representations
will appear in the OPEs of $\tPhi_\frac{1}{2}$ with other primary fields.
Since operators in the other indecomposable representations  are
logarithmic operators, this means that if $\tPhi_\frac{1}{2}$ is part of
the spectrum of the WZNW model, there will also be logarithmic operators
in the spectrum. This is the same situation as is familiar for the 
pre-logarithmic puncture-type operators in the $c_{p,q}$ LCFTs, 
except
that in this case, we can see that it must occur
because the pre-logarithmic $\tPhi_\frac{1}{2}$ operator is already in an
indecomposable representation of the full symmetry algebra of the model.

The $\ln x$ term in eq. (\ref{logxz})
does not vanish for $j=1/2$, and so it will still be a logarithmic
operator of the CFT on the boundary. This appears to contradict the result
of \cite{kogan}, where it was found that for the singleton with mass
$m^2=-1$, which corresponds to the vertex operators  with $j=1/2$, all the
logarithmic correlation functions on the boundary become null. However,
we can check that there is no real contradiction, because 
the dipole fields with $m^2=-1$ considered in \cite{kogan} and \cite{gka}
actually have different actions and equations of motion, so provided
$\tPhi_j$ turns out to be the boundary-bulk Green function for the action
considered in \cite{gka}, it will indeed couple to a logarithmic operator
on the boundary.

In both \cite{gka} and \cite{kogan}, the action has the form of eq.
(\ref{dipoleaction}), but in \cite{kogan} $\mu$ is taken to be a constant,
$\mu^2=1$, while in \cite{gka} $\mu$ is taken to be given by $\mu^2 =
4j-2$ (in 3 dimensions). When $\mu \neq 0$ the two actions are equivalent,
since $\mu^2$ can always be set equal to $1$ by replacing $A$ and $B$ with
$A' = \mu A$ and $B' = \mu^{-1}B$.   When $j=1/2$ 
the two versions of the action are inequivalent, and if 
$\mu\neq 0$ the result of \cite{kogan} that there are no logarithms
on the boundary will apply. It is therefore more useful for us to take
$\mu^2 = 4j-2$ as in \cite{gka}, so that when $j=1/2$,
$\mu=0$, 
 and the action (\ref{dipoleaction}) reduces to 
\beq S = \int d^3x
\sqrt{g}\left(g^{\mu\nu}\partial_{\mu}A\partial_{\nu}B  +   AB \right) 
\label{m-1action}\eeq
and the equations of motion are simply
\beq
(\nabla^2+1)B = (\nabla^2+1)A = 0
\eeq
It was observed in \cite{l} that in the $AdS_3$/CFT correspondence, a
logarithmic operator on the boundary with dimension $1$,
corresponding to $m^2=-1$, would have equations of motion in the bulk
with no diagonal terms, but that was interpreted as meaning that there
could be no logarithmic operators on the boundary with $m^2=-1$.  To
see why this is not necessarily true, we can see what happens when we
start from the equations which determine the Green functions for $j \neq
1/2$, and take $j \goto 1/2$. The Green functions $K_{ij}$  are determined
by the equations \bea \left(\nabla^2 - m^2\right) K_{AA}  - \mu^2K_{BA} =
0, ~~~~~~~~  \left(\nabla^2 - m^2\right) K_{BB} =  0   \nonumber \\
\left(\nabla^2 - m^2\right) K_{AB} - \mu^2 K_{BB} =  0, ~~~~~~~~
\left(\nabla^2 - m^2\right) K_{BA}  = 0  \label{Kij}\eea together with the
boundary condition that $K_{ij}=0$ on the boundary of $AdS_3$. Even when
$j \neq 1/2$, these equations have more than one solution, depending on
whether we take $K_{ij}$ to be symmetric or not \cite{kogan}. If $K_{ij}$
is symmetric, the solution is  \beq
K_{BB}=0,~~~~~K_{AB}=K_{BA}=K,~~~~~K_{AA}=\frac{1}{2}\frac{dK}{dj} \eeq
where $K$ is the boundary-bulk Green function for a massive scalar field,
which was found in \cite{witten}, and in the
$(\phi,\gamma,\bar\gamma;x,\xb)$ coordinates is $\Phi_j$. The other
independent solution is
\beq
K_{BA}=0,~~~~~K_{AA}=K_{BB}=K,~~~~~K_{AB}=\frac{1}{2}\frac{dK}{dj}
\eeq
Although there are two choices for the Green functions they both lead to
the same two point functions on the boundary. The only differences is that
in the symmetric case, the field $A$ couples to the logarithmic operator
$D$ on the boundary and $B$ couples to the primary operator $C$, while in
the non-symmetric case, $A$ couples to $C$ and $B$ to $D$ \cite{kogan}. 
If we choose the normalization of \cite{gka}, both solutions still apply
when $j=1/2$, because then $\frac{dm^2}{dj}=0$ and so $\frac{d}{dj}$
commutes with $(\nabla^2-m^2)$. The calculation of the two point functions
using (\ref{adscft}) then proceeds in exactly the same way when $j=1/2$ as
when $j \neq 1/2$, so either choice still leads to logarithms  on the
boundary. The difference is that there is now a third solution for the
Green functions, which is just $K_{AA}=K_{BB}=K$, $K_{AB}=K_{BA}$ =0.
Since the Green functions are then the same as for ordinary scalar fields,
there will be no logarithms on the boundary if we make this choice. We
therefore cannot determine if there will be logarithmic operators on the
boundary just using the action (\ref{m-1action}) -- we need some
information about interactions in the theory. Instead, we
can try to determine whether the spectrum of the CFT on the world-sheet
includes $\tPhi_{\frac12}$, which would indicate that we do have a
logarithmic operator on the boundary, or only $\Phi_{\frac12}$ in which
case there are no logarithmic operators.

\section{Conformal Blocks and OPEs}

To decide if there will be logarithmic operators in the theory, we need to
know if only the operator $\Phi_{\frac12}$ occurs, or if $\tPhi_{\frac12}$
appears as well. We are therefore interested in the OPEs of primary fields
for which $\Phi_{\frac12}$, and possibly also $\tPhi_{\frac12}$ occur. If
there are no logarithmic operators, the leading terms in the OPEs are 
\beq
\Phi_{j_1}(x_1,z_1)\Phi_{j_2}(x_2,z_2 ) \sim 
\sum_{j_3}
\frac{C(j_1,j_2,j_3)}{|x_{12}|^{2(j_1+j_2-j_3)}
|z_{12}|^{2(h_1+h_2-h_3)}} \Phi_{j_3}(x_2,z_2) 
\label{OPE1}\eeq
In general there is also another contribution to the OPE which is an
integral over $j=1/2+is$, which we ignore here. If all the OPEs are of
this form, there can be no logarithms in correlation functions. The other
possibility is that, when $j_3=1/2$ appears in the OPE, it becomes
\bea
\Phi_{j_1}(x_1,z_1)\Phi_{j_2}(x_2,z_2 ) &\sim&
\frac{C(j_1,j_2,\frac12)}{|x_{12}|^{2(j_1+j_2-j_3)}
|z_{12}|^{2(h_1+h_2-h_3)}}
\left[\ln|x_{12}|^2\Phi_{\frac12}(x_2,z_2) + \tPhi_{\frac12}(x_2,z_2)
 \right]
\nonumber \\
&&+~\sum_{j_3}
\frac{C(j_1,j_2,j_3)}{|x_{12}|^{2(j_1+j_2-j_3)}
|z_{12}|^{2(h_1+h_2-h_3)}} \Phi_{j_3}(x_2,z_2) 
\label{OPE2}\eea
In either case the OPE also has descendant terms of higher order in $x$
and $z$.  The possible values of
$j_3$ which occur in the OPE  can be determined from the three point
functions of primary fields which were calculated in \cite{teschner,ios}.
They are give by the poles of the function (from eq. (5.29) of \cite{ios})
\beq
D(j_a) \sim
\left(\frac{1}{k}b^{-2b^2}\frac{\Gamma(b^2)}{\Gamma(1-b^2)}
\right)^{1-j_1-j_2-j_3}\frac{1}{\Upsilon((j_1+j_2+j_3-1)b)} 
\sum_{a=1}^3\frac{\Upsilon(2j_ab)}{\Upsilon((j_1+j_2+j_3-2j_a)b)}
\eeq 
where $b^2=1/(k-2)$ and $\Upsilon(x)$ is the $\Upsilon$-functions defined
in \cite{zz}. $\Upsilon(x)$ has zeros at
\beq
x=-mb-\frac{n}{b},~~~~~(m+1)b+\frac{n+1}{b},~~~~~~ 
m,n \in {\bf Z}_{\geq 0}
\eeq
and so $D(j_1,j_2,j_3)$ has poles when one of the combinations
$j_1+j_2-j_3$, $j_2+j_3-j_1$, $j_3+j_1-j_2$ or $j_1+j_2+j_3-1$ takes one
of the values $(m+1)+(n+1)(k-2)$ or $-m-n(k-2)$, for a pair of
non-negative integers $(m,n)$. To have $j_3=1/2$, we can therefore choose
$j_1,j_2$ so that $2(j_1-j_2) \in {\bf Z}$. However, we cannot tell from
the three-point function which type of OPE we have, because both
(\ref{OPE1}) and (\ref{OPE2}) lead to the same three point function for
$\langle\Phi_{j_1}\Phi_{j_2}\Phi_{\frac12}\rangle$. The logarithmic term in
eq. (\ref{OPE2}) does not contribute because the two point function of
$\Phi_{\frac12}$ must be null, as in eq. (\ref{logxz}), if
$\tPhi_{\frac12}$ is in the OPE. To decide which OPE is correct, we can
consider the four point functions
\beq
\langle \Phi_{j_1}(x_1,z_1) \Phi_{j_2}(x_2,z_2) \Phi_{j_1}(x_3,z_3)
\Phi_{j_2}(x_4,z_4)
 \rangle
 \eeq
which can be written in the form of eqs. (\ref{4point},\ref{blocks}) with
$j_3=j_1$ and $j_4=j_2$. Whichever OPE is correct (provided there are no
$\ln z$ terms), the s-channel conformal block which contains the
contribution of $\Phi_J$ to the OPE of $\Phi_{j_1}$ and $\Phi_{j_2}$ can be
written as 
\beq 
F_J(x,z) = z^{h_J-h_1-h_2}\sum_{n=0}^\infty z^nF^n_J(x)
\eeq
The function $F^n_J(x)$ then contains all the contributions to the OPE of
$\Phi_{j_1}\Phi_{j_2}$ which are descendants of $\Phi_J$ of Virasoro level
$n$. In particular, $F^0_J(x)$ determines the OPE coefficients of all the
operators $(J^-_0)^n\Phi_J$. The KZ equation (\ref{KZ}) becomes a
system of equations for $F^n_J(x)$: 
\beq \left[ {\cal P}
-(k-2)(h_J-h_1-h_2-n)\right] F^n_J =  \left[ {\cal P} + {\cal Q} -
(k-2)(h_J-h_1-h_2-n+1) \right]F^{n-1}_J \label{FJn}
\eeq 
 where $\cal P$
and $\cal Q$ are given by eq. ({\ref{PQ}) with $j_3=j_1$ and $j_4=j_2$. If
we now write $F^0_J(x) = x^{J-j_+}\tilde{F}^0_J$, $j_\pm=j_1 \pm j_2$, the
equation for $\tilde{F}^0_J$   becomes a
hypergeometric equation and the solution is 
\beq
F^0_J(x) = f_J(x) \equiv x^{J-j_+}F(J+j_{-},J-j_{-};2J;x)
\label{fj}\eeq
The other solution of the equation for $F^0_J$ is $f_{1-J}(x)$, provided
$2J$ is not an integer, so  in order to have an OPE which only includes
operators $\Phi_J$ with $J \geq 1/2$, we have to choose the first
solution. If $J \geq 1$ and $2J \in {\bf Z}$, the second solution
instead has the form
\beq
f^{(2)}_J(x) = \ln x f_J(x) + x^{1-J-j_+}H(x)
\eeq
where $H(x)$ is a function which is regular at $x=0$. A four point
function of this form would imply both that the OPE included a primary
field with spin $1-J$ which is less than $1/2$, and that the OPE has a
$\ln x$ term but no $\ln z$ term, which from the previous section we know
can only happen for $J=1/2$, so we again have to take the first solution. 
If $j_1=j_2$, the identity operator with $J=0$ contributes to
the s-channel expansion and we have
\beq
f_0(x) = x^{-2j_1}F(0,0;0;x) = x^{-2j_1}
\eeq
In this case $F^0_{J=0}$ has only a single term, since the operator
$\Phi_0$ in the OPE is just the identity, and not the non-trivial operator
with $J=0$ we considered in section (2), so $J^a_0\Phi_0=0$ and does not
appear in the OPE. The second solution in this case is $f_1(x)$, which
gives the conformal block for $J=1$.
In all cases, Eq. (\ref{FJn}) can be turned into a set of algebraic
recursion relations using 
\bea
{\cal P} f_J(x) &=& \left[ (J(1-J) - j_1(1-j_1) - j_2(1-j_2) \right]f_J(x)
\nonumber \\
\left[{\cal P + \cal Q}\right] f_J(x) &=& \left[-x(1-x) \frac{d^2}{dx^2} +
(2x-1)\frac{d}{dx}\right]f_J(x) \nonumber \\
&=& \sum_{J'=J,J\pm 1}C^{J'}_{J} f_{J'}(x)
\eea
where
\bea
C_J^{J-1} &=& -(J-j_+)^2 \nonumber \\
C_J^J &=& -\frac12 \left[ J(J-1)^2 - j_+(j_+-2) - j_-^2 +
\frac{(j_+-1)^2j_-^2}{J(J-1)}\right]
  \\ 
C_J^{J+1} &=&  \left[\frac{(j_-+J)(j_--J)}{4J^2(2J+1)(2J-1)} \right]
\nonumber \\ && \times \left[ 
J^2((J-1)^2 + 2j_+(J-1) +j_+^2-j_-^2) + 2Jj_-^2(1-j_+) - j_-^2(j_+-1)^2
\right] \nonumber
\eea
Thus, $F_J^n(x)$ can be expressed as a sum of the functions $f_K(x)$, with
$K = J,J\pm 1, \cdots, J\pm n$, for any $J \neq 1/2$, and, as we expected,
the OPE has  the form of eq. (\ref{OPE1}) with no logarithmic operators.
The one remaining case to consider is $J=1/2$. This can occur if $2j_- \in
{\bf Z}$, so the simplest possibility is to take $j_-=1/2$. Then we have
\beq
f_\frac12 = x^{\frac12 - j_+}F(1,0;1;x) = x^{\frac12 - j_+}
\eeq
This will give us a conformal block with no logarithms,
but a solution of this form implies that $\Phi_\frac12$
appears in the OPE but $J^-_0\Phi_\frac12$ does not, so that the three
point functions satisfy
 \bea
\langle \Phi_{j_1}(x_1,z_1)\Phi_{j_2}(x_2,z_2)\Phi_\frac12(x_3,z_3)
\rangle &\neq& 0  \\
\langle
\Phi_{j_1}(x_1,z_1)\Phi_{j_2}(x_2,z_2)
\left[J^-_0\Phi_\frac12(x_3,z_3)\right] \rangle &=& -
\frac{\partial}{\partial x_3} \langle
\Phi_{j_1}(x_1,z_1)\Phi_{j_2}(x_2,z_2)\Phi_\frac12(x_3,z_3)  \rangle = 0
\nonumber
\eea
These cannot both be true, so the only way to get an OPE with no $\ln x$
terms is if $\Phi_\frac12$ does not appear at all. To get an acceptable
OPE including $\Phi_\frac12$ we therefore have to take the  general
solution for $F_{\frac12}^0$, which  is  
\beq
F_{\frac12}^0(x) = x^{\frac12 - j_+} \left[ A\ln\left(\frac{x}{1-x}\right)
+B \right]
\label{logcb}\eeq
and the recursion relations for $F_{\frac12}^n$ are then
\beq
\left[ {\cal P} - \frac{1}{8} + \frac{j_+^2}{2} -j_+     +n(k-2) 
\right] F^n_\frac12 =\left[ {\cal P} + {\cal Q} - \frac{1}{8} +
\frac{j_+^2}{2} -j_+ +  (n-1)(k-2)
\right]F^{n-1}_\frac12
\eeq
A conformal block of the form (\ref{logcb}) implies that the leading
term in the OPE must have the form of eq. (\ref{OPE2}), and so we can see
that it is impossible to have $\Phi_\frac12$ in the spectrum of the WZNW
model without $\tPhi_\frac12$ also being included.

Finally, we can show that the inclusion of $\tPhi_\frac12$ also implies
that the spectrum must include other logarithmic operators, which will
have correlation functions which depend on  $\ln z$ as well as $\ln x$, by
considering the OPE of $\tPhi_\frac12$ and $\Phi_j$. The form of the OPE
 for
$\Phi_\frac12\Phi_j$ will as usual be given by eq. (\ref{OPE1}),
and then the OPE of $\tPhi_\frac12\Phi_j$ can be found by applying the
currents $J^a_0$ to both sides of the OPE, or more simply by
differentiating eq. (\ref{OPE1}) wrt $j_1$, giving
\beq
\tPhi_\frac12(x_1)\Phi_j(x_2) = 
\sum_{j_3}
\frac{C(\frac12,j,j_3)}{|x_{12}|^{2(1+j-j_3)}
|z_{12}|^{2(h_\frac12+h_j-h_3)}} \left[ 
\ln |x_{12}|^2\Phi_{j_3}(x_2,z_2) + \tPhi_{j_3}(x_2,z_2) \right]
\eeq
The only way to avoid having logarithms in both $x$ and $z$ in the WZNW
model is therefore if all correlation functions of the primary field with
$j=1/2$ vanish.

\section{Conclusions}

We have shown that the CFTs on the world-sheet and the boundary of $AdS_3$
can both have logarithmic operators. We therefore have a duality between
two LCFTs. In most cases, primary operators on the world sheet are mapped
to primary operators on the boundary and logarithmic operators are mapped
to logarithmic operators. However, because logarithmic operators on the
boundary have to be in indecomposable representations of the global part
of the Kac-Moody algebra on the world-sheet, but do not have to be in
indecomposable representations  of the world-sheet Virasoro algebra, it is
possible for primary operators in one CFT to correspond to logarithmic
operators in the other. The field with $j=1/2$ is one example -- it is a
logarithmic operator on the boundary but not on the world-sheet. An
operator which was in an indecomposable of the world-sheet Virasoro
algebra but not of the global $SL(2)$ algebra would be logarithmic on the
world-sheet but not on the boundary -- the logarithmic operators which
occur in finite dimensional representations of $SL(2)$ are of this type.
It is therefore possible that in other models a LCFT could be dual to an
ordinary CFT -- for example, there could be logarithmic operators in the
world-sheet CFT for string theory on $AdS_{D>3}$ which describe d-brane
recoil in those theories, without there being logarithmic operators in the
CFT on the boundary. We have also seen that most, but not all, of the
logarithmic operators in either CFT correspond to dipole fields in the
bulk.

The fields with $j=1/2$ play a crucial role in determining whether 
 the WZNW model on $SL(2,R)$ or $SL(2,C)/SU(2)$ is a LCFT or not.
 This is the representation which has the minimum value of the
conformal dimension in the discrete series, and the maximum value in the
continuous series. It also corresponds to the minimum mass$^2$ for scalar
fields in $AdS_3$, it is the only field which can be in an indecomposable
representation of $SL(2)$ but still be a primary field in an irreducible
representation of the Virasoro algebra on the world sheet, and it is the
only field which can couple to a logarithmic operator on the boundary of
$AdS_3$ even if it has the same equation of motion in the bulk as an
ordinary scalar field. Like the puncture operator in minimal models, the
operator $\tPhi_\frac12$ is not a logarithmic operator (on the
world-sheet) but logarithmic operators occur in the OPE of 
$\tPhi_\frac12$ with other primary fields. However, unlike the puncture
operator, $\tPhi_\frac12$ is not in an irreducible representation of the
complete symmetry algebra of the theory, and this provides a simple way to
understand why the logarithmic operators must be included in a theory with
$\tPhi_\frac12$ in the spectrum.

The
major question raised by these results is, what are the
implications for string theory on $AdS_3$?
Logarithmic operators in string theory are thought to generate additional
target space symmetries, and it would be interesting to know what zero
modes in the string theory are generated by the logarithmic operators in
this WZNW model-- for instance, it has been suggested that logarithmic
operators might  restore the Poincare symmetry broken by the position of
the branes in the D1-D5 system.   It is therefore important to find out
for which values of $j$ the logarithmic operator as well as the primary
operator is part of the spectrum. The simplest solution  would be if the
fields with $j=1/2$ decoupled completely from the spectrum, in which case
there would be no logarithmic operators, but this seems to conflict with
the 3-point functions calculated in \cite{teschner,ios}. 
We have found that  if there is a scalar field with
$m^2=-1$ in the bulk theory, there must also be dipole fields with
$m^2>-1$. It would be interesting to understand why this should be true
in the supergravity theory.

\end{document}